\documentstyle[11pt,paspconf]{article}

\begin{document}

\title{The Color-Magnitude Relation of Early-Type Galaxies: 
A Tool for Cluster Finding and Redshift Determination}
\author{H.K.C. Yee, M.D. Gladders}
\affil{Department of Astronomy, University of Toronto, Toronto, 
ON, M5S 3H8, Canada}
\author{O. L\'opez-Cruz}
\affil{INAOE-Tonantzintla, Puebla, M\'exico}

\begin{abstract}
The color-magnitude relation (CMR) of early-type galaxies in
  clusters
provides a powerful tool for finding clusters and determining
photometric redshifts for clusters using two-filter imaging data.
We demonstrate the high accuracy of photometric redshifts
attainable using the CMR by applying the technique to a sample of 45
Abell clusters with known redshifts.
Furthermore, by using the red sequence of galaxies from the CMR, 
we have developed an extremely efficient technique for detecting
clusters and groups of galaxies.
We demonstrate this using both observed photometric catalogs with redshifts
(from the CNOC2 Survey) and simulations.
\end{abstract}


\section{Introduction}
 Early-type galaxies form the dominant population in the core of all
galaxy clusters with few or no exceptions---from low redshift to $z>1$,
from poor to rich clusters.
The uniformity of the properties of early-type galaxies provide 
important tools, such as the fundamental plane and the color-magnitude
relation (CMR), in the study of galaxy clusters and their evolution.

  The CMR for early-type galaxies (E/S0's) was first noted by Baum (1959)
in that the colors for elliptical galaxies become bluer as they
become less luminous.
Furthermore, E/S0 galaxies are also the reddest normal galaxies at any
single given redshift.
Hence, the E/S0's form a sequence on the color-magnitude plane which
is sometimes called the ``red sequence'', as illustrated for the cluster
Abell 2256 using data from L\'opez-Cruz \& Yee (1999a, also see L\'opez-Cruz
1997) in Figure 1.
The existence of the CMR is in general attributed to a metallicity
effect due to the varying ability of spheroids of different masses to retain
metals in the presence of supernovae driven winds (e.g., see Kodama
\& Arimoto 1997 and references therein).
The dispersion, color, and slope of the red sequence in clusters have
been used to study the evolution of cluster galaxies up to a redshift
as high as $\sim 1$ (e.g., Bower et al. 1992; Arag\'on-Salamanca et al. 1993;
Stanford et al. 1998; and Gladders et al. 1998).

In this paper, we demonstrate the efficiency of using the color-magnitude
relation of clusters as a powerful tool for estimating the redshift of 
clusters photometrically and detecting clusters from two-filter 
imaging data.

\begin{figure}
\plotfiddle{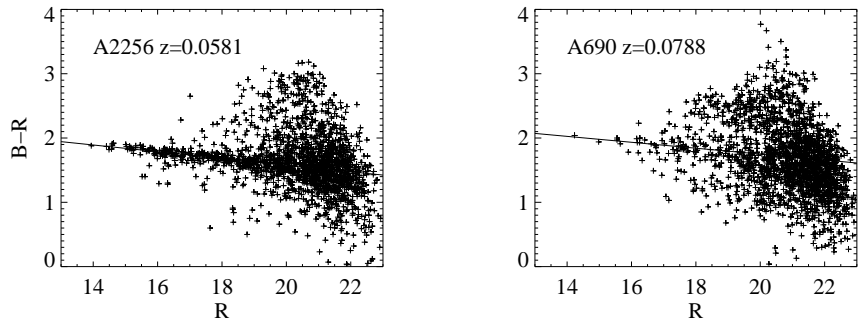}{4.0cm}{0}{70}{70}{-190}{-12}
\caption{Color-magnitude diagrams for Abell 2256 and Abell 690, with
the best CMR fits indicated by solid lines.}
\end{figure}

\section{The CMR as a Redshift Estimator}

The red sequence of early-type galaxies in clusters provides a powerful
means for estimating the cluster redshift using only two filters.
The efficacy of using the red sequence to determine redshift arises from
two factors.
First, early-type galaxy spectra, because of the large 4000\AA~break, provide
the largest signal for photometric redshift determination.
By predetermining the spectral type of the galaxies, (i.e., picking the
reddest galaxies available by virtue of the prominence of the red sequence),
only a single color (from two filters, assuming that they straddle the
4000\AA~break) is needed to provide the redshift
signature.
Second, the red sequence is composed of many such galaxies; hence,
if the whole sequence is used instead of a single galaxy, the
measurement uncertainty would diminish significantly.

To test the efficiency and accuracy of this method, we
use two-color CCD photometry from a sample of 45 Abell clusters from
the Low-Redshift Cluster Optical Survey (LOCOS) of L\'opez-Cruz \& Yee
(1999a).
The clusters have a redshift range between 0.025 to 0.18.
The LOCOS data set consists of $B$$R$$I$ photometry obtained using
the 2K$\times$2K T2KA CCD on the KPNO 0.9m over a field size of  
23$'\times$23$'$
to a depth of typically $M_R=-17.5$ ($H_0=50$ km/s/Mpc).
The details of the analysis of the CMR of the clusters are presented
in L\'opez-Cruz \& Yee (1999b) and Gladders et al. (1998).
Briefly, the CMR for each cluster is fitted with a linear function
on the $B-R$ vs $R$ color-magnitude
plane, using a robust regression method with outlier rejection.
The resultant slopes of the fits indicate  that the CMR is consistent
with being universal in the redshift range of the sample, reflecting
the high $z$ formation epoch of the red galaxies in the clusters
(see also Gladders et al. 1998).
We note that in order to apply the color of the red sequence to obtain
a photometric redshift, we must assume that the red galaxies in 
different clusters at the same $z$ are coeval -- a condition that 
is approximately met if the cores of clusters are formed at high $z$.

From the fit of the CMR, the $B-R$ color of a red sequence at any $R$ can
be determined.
We have chosen a fiducial, though arbitrary, magnitude of $R=17$ for
the purpose of obtaining an empirical calibration between $z$ and
the $B-R$ color.
We note that an apparent magnitude is chosen since in determining a
photometric redshift, no prior knowledge of the redshift can be assumed.
The correlation between $z$ and the $B-R$ color of the red sequence at
$R=17$ is shown in Figure 2.
A second degree polynomial fit to the data produces a residual 
in $z$ of 0.008, or a $\Delta z/z\sim0.1$.
Comparing to the residual of $\sim0.04$ to 0.07 ($\Delta z/z\sim 0.1$ to 0.2)
 of typical empirical photometric redshift techniques using four colors
(see e.g., Brunner et al. 1997, Yee 1999, and papers in these proceedings),
the empirical calibration of the cluster redshifts using a single CMR color
is remarkably tight and robust.

\begin{figure}
\plotfiddle{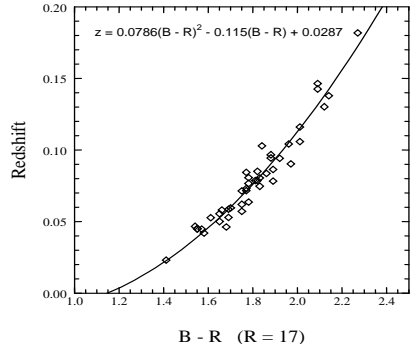}{3.9cm}{0}{35}{30}{-135}{-68}
\caption{Empirical calibration of CMR colors and $z$ for a sample
of low-$z$ Abell clusters} 
\end{figure}

\section{The CMR as a Tool for Detecting Galaxy Clusters}

The ubiquitousness of the early-type galaxy red sequence can be used
as a powerful tool for detecting high spatial density regions such
as galaxy clusters and groups.
Selecting galaxies in color slices in the color-magnitude diagram
would preferentially increase the density enhancement of any
grouping of early-type galaxies which are normally found
in clusters or groups.
This technique can be used not only as a very efficient cluster finding
algorithm, but also as a means of separating projected clusters and groups.

\subsection{Separating Projected Clusters}
Cluster catalogs created by primarily locating surface density enhancements
of galaxies, such as the Abell and APM catalogs, suffer from projection
effects.
Estimates of such projection effects in the Abell catalog range from 10 to
20\% (e.g., Olivier et al. 1990).
Figure 1 demonstrates the  ability of using the color-magnitude diagram
 of a cluster field
to detect background clusters in an efficient manner.
A second red sequence above the red sequence of A690 can be clearly
seen.
The galaxies in this second red sequence, which corresponds to a redshift
of $\sim$0.25, are found to be clustered
to the NW of the main A690 cluster (at $z=0.079$), and coincident 
with a secondary X-ray peak.

In the LOCOS data set, we identified 6 clusters out
of the 45 as having a significant background cluster 
(L\'opez-Cruz \& Yee 1999b),
giving a contamination rate of $\sim$10 to 15\% for the Abell clusters.
A similar rate (2/9) is also obtained for the control fields.
We note that such contaminations are also present in the X-ray maps,
although at a smaller rate of 1/2 to 1/3 of the optical images.

\subsection{A Red Sequence Cluster Survey}

A major problem in existing
galaxy cluster catalogs from optical surveys (e.g., Abell 1958, Postman
et al. 1996) is that they are susceptible to projection effects.
An additional difficulty for any survey that attempts to find high-redshift
clusters is the accumulated column density of galaxies in the foreground
of a cluster, which decreases the surface density enhancement of galaxies
in the cluster.
The red sequence of early-type galaxies in clusters and groups, however,
provides a crucial signature which allows us to identify such high-density
regions.

We are in the process of carrying out a large optical imaging survey
for galaxy clusters to redshifts $>\sim1$,
using the red sequence as a primary marker (Gladders \& Yee 1999).
The survey, conducted at the CFHT 3.6m and the 
CTIO 4m using large mosaic CCD imagers,
will obtain images over 100 deg $^2$ in area in the $z'$ ($\sim$9200\AA)
and $R$ bands  sufficiently deep to
detect galaxy clusters to $z\sim1.4$.
The main scientific goals are to obtain the mass spectrum of clusters
of galaxies as a function of redshift in order to constrain the
cosmological parameter pair: ($\Omega_m,\sigma_8$), and to study the
evolution of clusters over a large redshift range with a sample
of clusters covering a wide range of properties.

The cluster detection algorithm uses combined information on the 
color-magnitude plane and the x-y position plane.
Galaxies are chosen based on slices in the $z'-R$ vs $z'$ plane which
mimic the CMR of early-type galaxies at various redshifts.
Model CMRs as a function of redshift can be obtained from 
synthesis models from, e.g., Kodama (1997).
Surface enhancements of galaxy density in the x-y plane in these slices 
are then cataloged as cluster candidates.
For images with good seeing, morphological information in the form
of a concentration parameter can also be used to further isolate 
early-type galaxies.
In short, clusters are identified in the 5-dimensional space of RA, Dec,
color, magnitude, and morphology.

The use of the red sequence minimizes projection effects from fore- and
background clusters and groups, 
as the color allows a crude separation in redshift space.
Furthermore, by isolating one color slice at a time (and with 
added morphological selection when available), the excess signal from
the cluster is greatly enhanced.
The detected red sequence can then be used to estimate a photometric
redshift for the candidate, using either CMR models or an empirically
calibrated relation.
With a photometrically measured redshift, global properties such 
as the luminosity function, richness, and blue fraction for the cluster can be 
estimated statistically from the photometric data.

\medskip
\begin{figure}
\plotfiddle{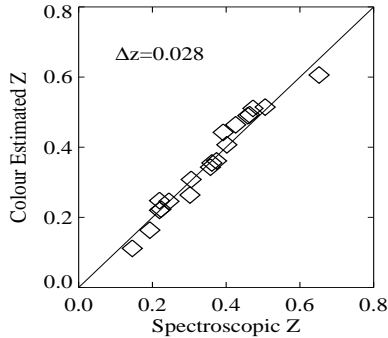}{3.8cm}{0}{60}{50}{-110}{-17}
\caption{Comparison of photometrically determined $z$ using the CMR
and spectroscopic $z$ for the clusters/groups found by the red sequence 
method from the CNOC2 database.}
\end{figure}
\noindent
{\it 3.2.1.~ Testing the Red Sequence Algorithm with CNOC2 Data}
\smallskip

\noindent
We have tested the cluster detection algorithm using both real and
simulated data.
The CNOC2 survey (Yee et al. 1998) provides 
the largest redshift/multi-color photometry database currently
available for intermediate redshifts.
Using 3/4 of the data (about 1.2  deg$^2$), we construct a catalog
of groups/clusters using $I$ and $g$ photometry and the method outlined
above.
For each of the identified groups, a photometric
redshift is determined by comparing with CMR models from Kodama (1997).
The groups are then verified using the redshift catalogs.
Every identified group is found to have a corresponding group in redshift
space.
The average richness of these groups is sub-Abell richness class 0, with
a mean velocity dispersion of about 350 km/s.
Figure 3 shows the comparison of the photometric redshifts as derived
from the red sequence with the spectroscopic redshifts of the groups.
The agreement is excellent over the redshift range of 0.1 to 0.7
 with a mean $\Delta z$ of only 0.028,
comparable to or better than the best photometric redshift  
results using 4 colors (e.g., Brunner et al. 1997).
This demonstrates that the red sequence method is able to detect very
poor clusters and groups and to estimate their redshifts with remarkably
high accuracy using CMR models.

\medskip

\noindent
{\it 3.2.2.~ Testing the Red Sequence Algorithm with Simulated Data}
\smallskip

\noindent
To test the efficiency, robustness, and systematic effects of the red sequence
method, we have also constructed detailed models of the observed sky projection
of field galaxies.
The simulations of the background field galaxy distribution reproduce
known galaxy count distributions (to AB magnitude $\sim 29$),
galaxy luminosity functions, galaxy redshift distributions, color
distribution, density morphology relation, and angular covariance function.
Clusters of different properties, such as blue fraction, richness, 
ellipticity, and core radius etc., at different redshifts are then placed
in the simulated field.
The cluster finding and photometric redshift procedures are then
applied to the simulated data.
Figure 4 shows the results of one set of simulations in which the blue
fraction is varied for a sample of Abell richness 1 clusters. 
The simulated galaxy photometry catalogs have the same depth as the
cluster survey with a 5$\sigma$ detection limit of $z'=23.6$ and $R=24.8$.
It is found that the detection probability essentially stays constant
at near 100\% for clusters with blue fractions less than 0.8 up to
$z\sim1.2$; and the detection probability remains significant at $\sim$50\%
at $z\sim1.4$.
At $z>1.4$, the $z'$ band passes blueward through the 4000\AA~break, 
essentially erasing the advantages of using the red sequence method.
Poorer clusters, of about Abell richness class 0, can 
be detected at near 100\% detection 
probability up to $z\sim$0.9.
Other tests indicate that properties such as the  ellipticity and core 
radius of the clusters have little or no effect on the detectability.

\begin{figure}
\plotfiddle{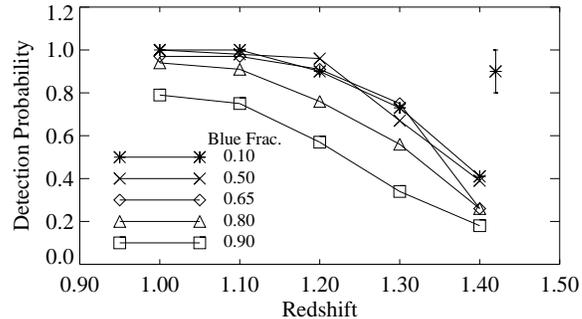}{3.7cm}{0}{70}{70}{-120}{-15}
\caption{Detection probabilities of simulated clusters of Abell 1 richness with
different blue galaxy fractions as a function of $z$.}
\end{figure}

\section{Summary}

The CMR is a powerful tool for detecting galaxy clusters and estimating
their redshifts.
We have demonstrated that at low-redshift where one can obtain an
empirically calibrated relation between the color of the red sequence
and redshift, cluster redshifts can be estimated to a high accuracy 
($\sim0.01$) using just two filters.
The red sequence is also an extremely efficient method for identifying
clusters over a large richness and redshift range.
We have begun a 100 deg$^2$ survey using modern mosaic CCD imagers on
4m class telescopes, from which we expect to find upward of 100
rich clusters at $z\sim1$.

\end{document}